\documentclass[aps,prl,twocolumn,superscriptaddress]{revtex4-1}
\usepackage{longtable}
\usepackage{amsfonts}
\usepackage{amsmath}
\usepackage{graphicx}
\usepackage{stackengine,graphicx}
\usepackage{epstopdf}
\usepackage[caption=false]{subfig}
\usepackage[colorlinks]{hyperref}

\begin{document}

\message{ !name(Localization in Suspended Graphene.tex) !offset(-3) }

 \title{Strong localization in a suspended monolayer graphene by intervalley scattering}

\author{Cenk Yanik}
\author{Vahid Sazgari}
\author{Abdulkadir Canatar}
\author{Yaser Vaheb}
\author{Ismet I. Kaya}

 \email{iikaya@sabanciuniv.edu}
 \affiliation{Faculty of Engineering and Natural Sciences, Sabanci University, Tuzla, 34956 Istanbul, Turkey}
 \affiliation{SUNUM, Sabanci University Nanotechnology Research and Application Center, Tuzla, 34956 Istanbul, Turkey}

\begin{abstract}

A gate induced insulating behavior at zero magnetic field is observed in a high mobility suspended monolayer graphene near the charge neutrality point. The graphene device initially cleaned by a current annealing technique was undergone a thermo-pressure cycle to allow short range impurities to be adsorbed directly on the ultra clean graphene surface. The adsorption process generated a strong temperature and electric field dependent behavior on the conductance of the graphene device. The conductance around the neutrality point is observed to be reduced from around $e^2/h$ at 30 K to $\sim0.01~e^2/h$ at 20 mK. A direct transition from insulator to quantum Hall conductor within $\approx0.4~T$ accompanied by broken-symmetry-induced $\nu=0,\pm1$ plateaux confirms the presence of intervalley scatterers.

\centering
\end{abstract}

\date{\today}

\maketitle
%\section{I. Introduction}

The nature of the conductivity at Dirac point has been debated since graphene's first isolation~\cite{1}. One of the most important applications of graphene would be in digital electronics if it could be made to have depletable conductance while maintaining its high mobility. However, in graphene the on/off resistance ratio is  hindered by  potential fluctuations generally attributed to unintentional doping where minimum conductance is limited by saturation of the average carrier density in the presence of so-called electron-hole puddles.  Even  ultra-clean high mobility suspended monolayer graphene samples have been observed to have a minimum conductivity~\cite{9,SuspendedUltrahighM,12,13,14} complying with the theoretical ballistic limit ($4e^2/\pi h$)~\cite{17,18}. On the other hand, insulating behavior around Dirac point has been observed in double-layer graphene heterostructures~\cite{16} or in top-gated graphene sheets on hBN substrates~\cite{Amet_PRL2013} by screening the charge puddles. In these observations, the insulating regime is mediated by domination of intervalley scattering induced by atomic-scale defects or local sublattice symmetry breaking due to the hBN substrate randomly oriented with respect to the graphene sheet. It should be noted that a perfect rotational alignment between graphene and hBN lattices opens a large gap in the graphene Dirac point manifested by an activated insulating behavior~\cite{hBN-Gap1}, otherwise, with a random orientaion, a semimetallic behavior is expected as it is for graphene~\cite{hBN-Gap1,hBN-semimetal}.

According to the scaling theory of localization, when the spatial symmetry of a two dimensional system is broken, its conductivity tends to zero. In presence of impurities that have potential range extending much longer than the lattice constant, symmetry is preserved and there is no mixing between K and K' points in the band structure of graphene. This leads to a positive correction to the conductivity and anti-localization is predicted~\cite{suzuura2002crossover,suzuura_anderson}. On the other hand, by the addition of short range imputirities the symmetry is broken and intervalley scattering is allowed. In general, there are two scattering mechanisms for Dirac fermions in graphene, intra-valley and inter-valley scattering. In the presence of long-range disorder potentials, as in the case of graphene on Si substrate, the electrons scatter in each of the two valleys without backscattering~\cite{Long_range_disorder1,Long_range_disorder2,Long_range_disorder3}. However, with short-range or strong long-range disorders~\cite{Strong_Long_Range_disorder}, e.g., in graphene on hBN or suspended graphene, the dominant scattering is inter-valley scattering which gives rise to backscattering and localization~\cite{AL1,AL2,suzuura2002crossover,suzuura_anderson,mccann2006weak,AL3,Short_Range_disorder2,Short_Range_disorder3,16,Amet_PRL2013,Short_Range_disorder1}.

Here we report the first time observation of an insulating behavior in a suspended monolayer graphene around its charge neutrality point at zero magnetic field. This peculiar behavior, characterized by highly temperature-dependent strong conductance fluctuations, is mediated by the valley symmetry breaking and attributed to the presence of short-range disorders. The inter-valley scattering length is estimated to be $l_{iv} \approx 0.1~\mu$m by  gate and temperature dependent measurements as well as the magnetotransport data.

\begin{figure}[t]
\begin{center}
  \includegraphics[width=0.38\textwidth]{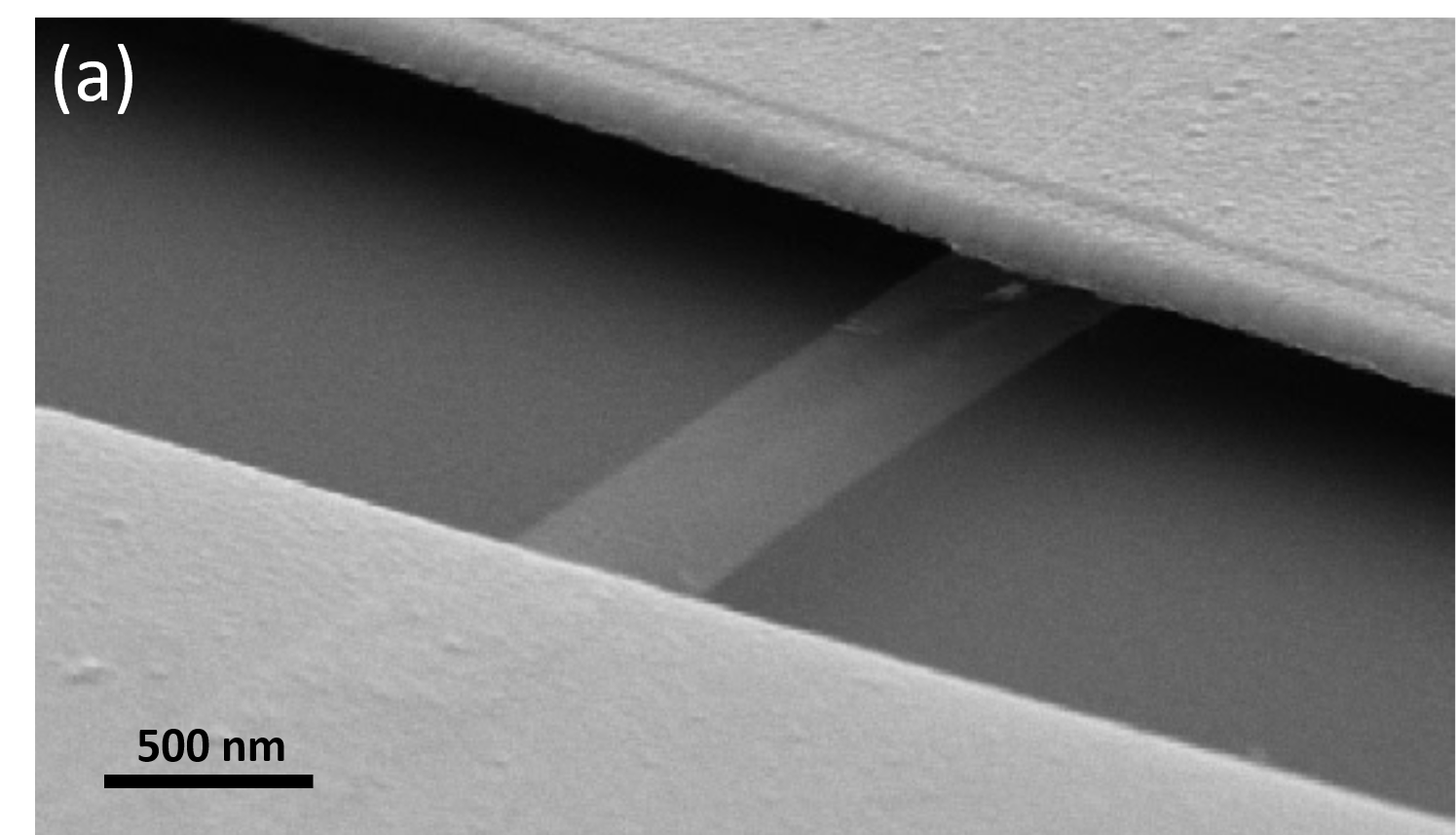}
  \includegraphics[width=0.5\textwidth]{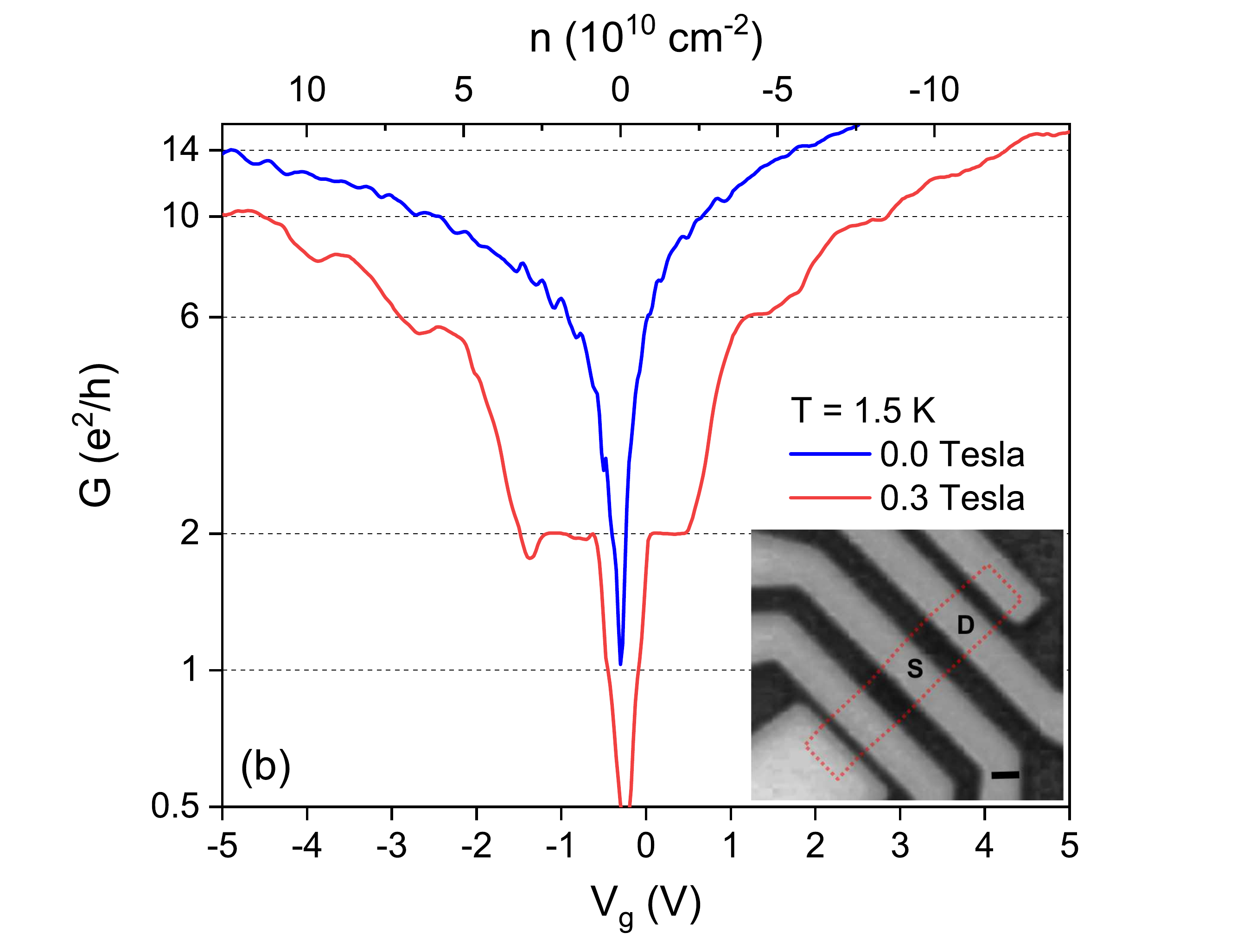}
\caption{The SEM image of one of the samples (not the one studied in this work) successfully suspended after BOE ething of the SiOx substrate. (b) The conductance at zero field (blue) and at  $B=0.3~T$ (red) as a function of carrier concentration, measured after current annealing of the suspended graphene but before it is underwent the thermo-pressure cycle. Device has channel length $L=1~\mu$m and width $W=2~\mu$m. The lead resistance $R_C=900 \pm 100~\Omega$ is estimated from the quantum Hall plateaus and subtracted in the plots. Inset shows the optical microscope image of the measured device. Measurements were performed between the probes labelled as S and D with $I_S=10~$nA applied current at $1.5$~K. Dashed lines mark the borders of the suspended graphene. Scale bar is $1~\mu$m.}\label{G-vs-V-beforTPC}
\end{center}
\end{figure}

%\section{II. Sample Preparation}
The suspended graphene sample was treated by a two-step procedure that involved removal of long range scatterers followed by deposition of short range scatterers. The sample was first cleaned by a current annealing  scheme, through which the graphene sheet and the contact probes were annealed concurrently~\cite{CurrentAnnScheme}, to the point that a very sharp conductance dip is obtained (Fig.~\ref{G-vs-V-beforTPC}(b)). Organics and residues left on graphene are known to generate long range density fluctuations in the form of electron-hole puddles which effectively saturate average carrier density and making the Dirac point unaccessible. In conventional current annealing method~\cite{moser2007}, the metal probes anchored to the cryogenic temperatures sink the current-induced heating power to the cooling system thus producing a highly non-uniform temperature profile on the graphene sheet. The temperature of the graphene can not be sufficiently high specially near the contact probes to globally remove the residual contaminations. However, the simultaneous annealing of probe and graphene helps to achieve a uniformly high temperature profile over a graphene sheet in a vacuum chamber. By application of huge currents through the metal probes which are narrowed near the graphene contact area, we managed to substantially elevate the temperature of the graphene near the metal contact area while the substrate maintained at cryogenic temperature of the cryostat. It improves the uniformity of graphene temperature compared to conventional single current annealing technique. This allows a thorough cleaning of graphene from  the contamination stuck on it before and during the fabrication process.  

In the second step the sample was let undergo a thermal cycle which also caused a brief and mild loosening of the vacuum level in the chamber.  An insulating behavior was acquired after the thermo-pressure cycle (TPC) of the high quality ultra-clean suspended graphene sample. In addition to the normal sequence of quantum Hall plateaus for single layer graphene, magnetoresistance measurements reveal emergence of indisputable $\nu =0,\pm1$ plateaus as a result of broken valley and spin symmetries~\cite{13}. This is interpreted to be due to the presence of strong short range scatterers that break the valley symmetry in an ultra-clean graphene sheet.

The graphene sample used in these experiments was mechanically exfoliated from a natural graphite and then transferred on to a $p$ doped Si substrate covered by 285 nm of SiO\textsubscript{x}. Single-layer flakes were identified based on their contrast under the optical microscope and confirmed by Raman spectroscopy. Electron beam lithography is employed to pattern the electrical contacts made from Cr/Au (3/100 nm) followed by a lift-off in acetone. Suspension is achieved by dipping the SiO\textsubscript{x} in a buffered oxide etcher (BOE) to remove $185$ nm of SiO\textsubscript{x} layer. This etchant offers a very controlled etching process with an etch rate
of about 1.2 nm/s. %The remaining oxide is allowed to avoid any shorts between the leads and Si layer.
Subsequently the device was transferred into DI water and isopropyl alcohol followed by a gentle nitrogen dry. Fig.~\ref{G-vs-V-beforTPC}(a) displays an SEM image of a typical sample we managed to suspend. A uniform etching of the SiOx layer underneath the graphene flake is clearly seen in this picture. We did not take SEM image of the studied sample because it would strongly affect or even damage the suspended flake.

Electrical measurements were done in a dilution refrigerator with a magnet using standard lock-in techniques. 
%{\color{red} Figure ??? shows the image of a suspended graphene sample fabricated by this technique which is not used for the measurements presented here.} 
The sample and the metallic leads were annealed at 1.5~Kelvin by passing independently controlled DC currents through them. This technique allowed heating of both graphene and leads individually to sufficiently high temperatures and prevented accumulation of residues near the leads. The details of the annealing  procedure is provided in Ref.~\cite{CurrentAnnScheme}. The annealing is done in repetitive current ramps with increased max current until the resistance peak shifted to near zero gate bias indicating low unintentional doping.

\begin{figure}[t]
\begin{center}
  \includegraphics[width=0.5\textwidth]{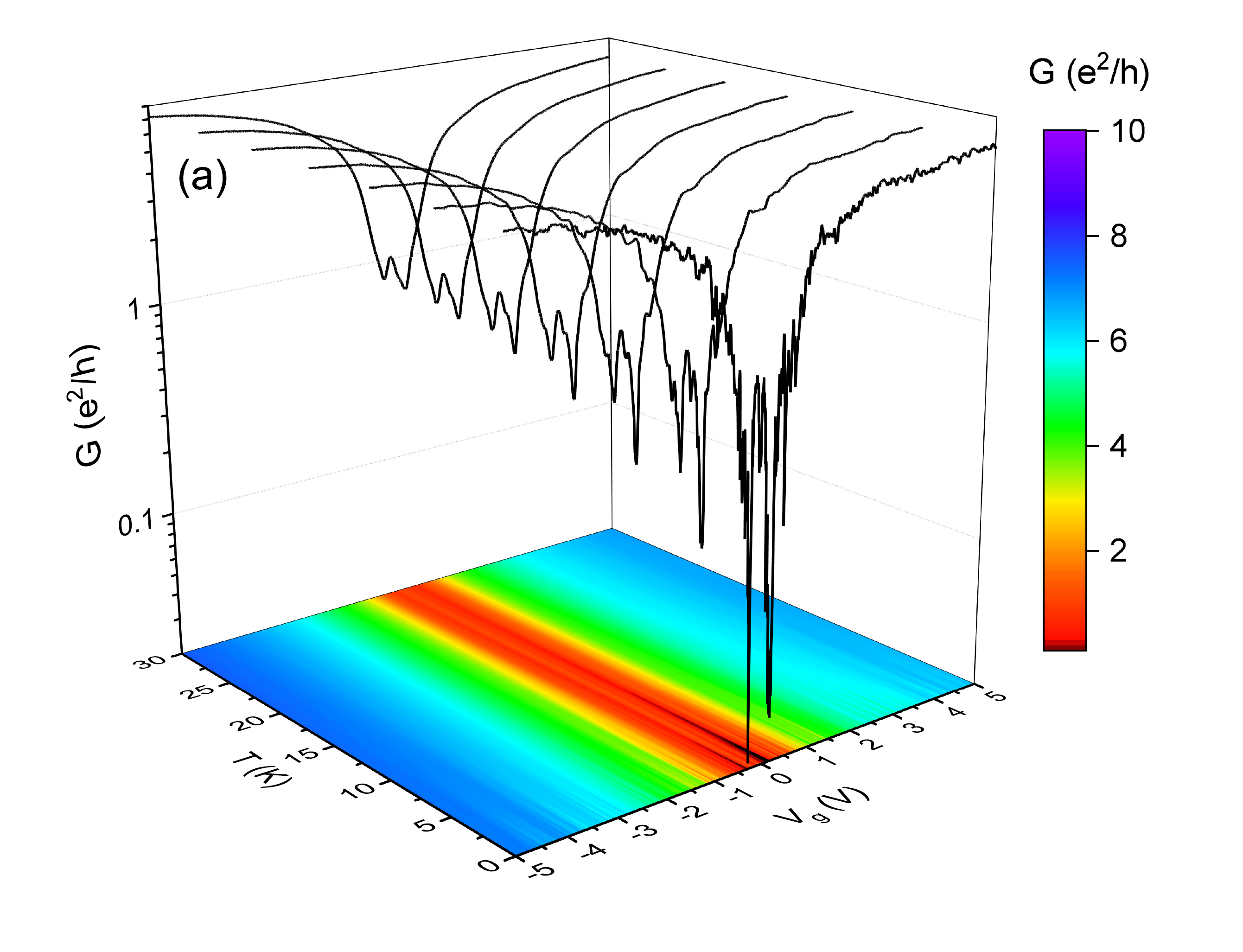}
   \includegraphics[width=0.5\textwidth]{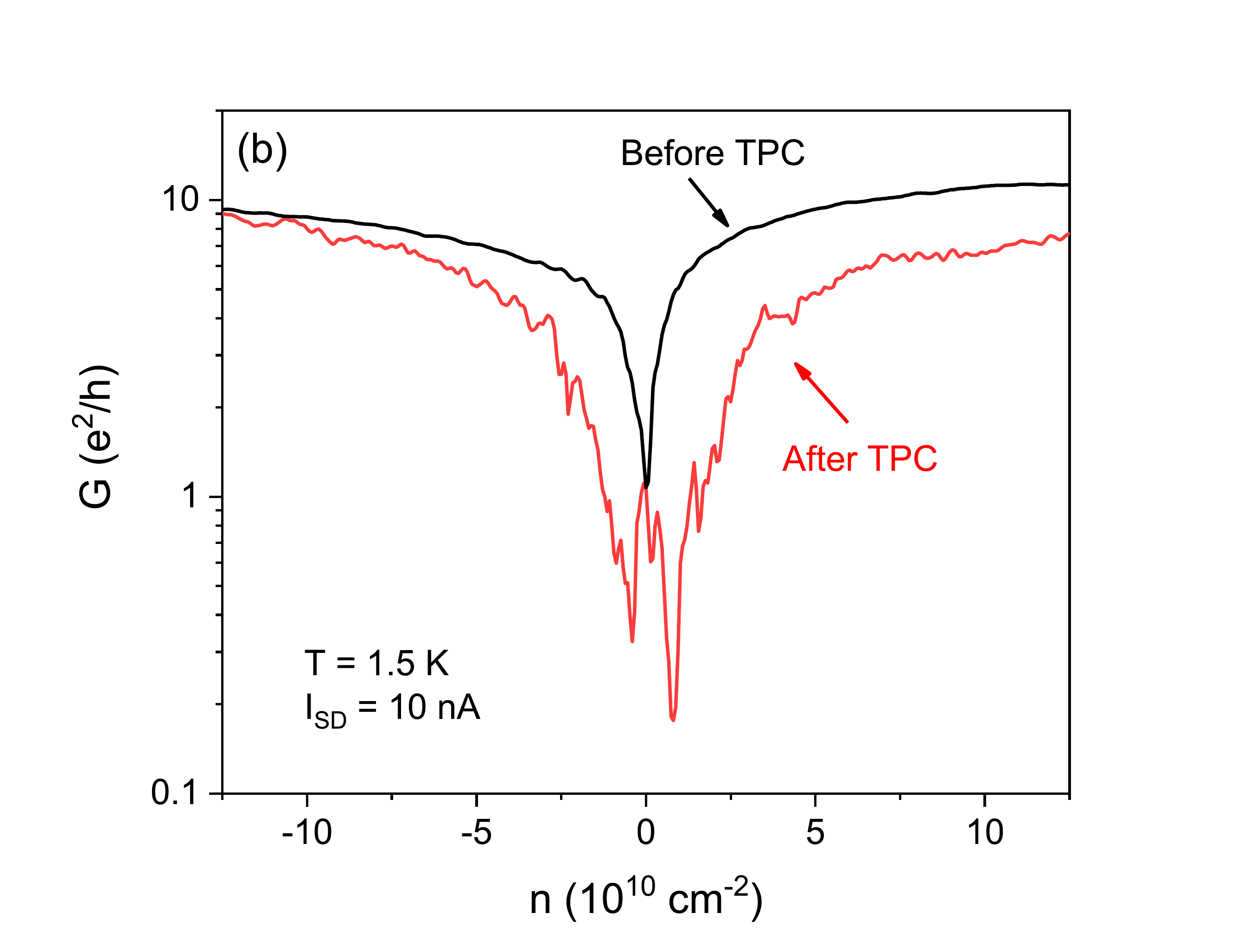}
\caption{(a) Conductance as a function of $V_g$ after thermo-pressure cycle at various temperatures at zero magnetic field. The insulating behavior appeared after  adsorption of short range impurities. (b) Conductance versus charge density is compared before and after the thermo-pressure cycle.}\label{G-V-T}
\end{center}
\end{figure}

The conductance of the sample after current annealing is displayed as a function of the gate voltage, $V_g$ and the carrier concentration, $n$ in Fig.~\ref{G-vs-V-beforTPC}(b). Parallel-plate capacitor model is used to determine the variation of $n$ with respect to the gate voltage as $n(V_g)=\alpha V_g$ where the value of the coupling factor is determined as $\alpha=2.7\times10^{10}$~V$^{-1}$cm$^{-2}$; this value  is consistent with the one estimated obtained from the quantum Hall (QH) measurements. 

%Carrier concentration varying by gate voltage, $n(V_g)=\alpha V_g$, is calculated with the coupling factor of $\alpha=2.7\times10^{10}~cm^{-2}$ obtained from the quantum Hall (QH) measurements in agreement with the parallel-plate capacitor model.

As shown in Fig.~\ref{G-vs-V-beforTPC}(b), the sample was confirmed to be a monolayer graphene via QH measurements where conductance exhibits well developed quantized plateaus at $\nu=\pm2,\pm6,\pm10$ at a magnetic field as small as $0.3$~T. Taking the aspect ratio of $W/L=2$ into consideration, it should be noted that the peak resistivity of the suspended graphene sample after current annealing ($\sim 50~k\Omega$) is well above the resistance quantum $h/e^2$ which is a hallmark of extremely clean samples with substantially reduced electron-hole puddles~\cite{Sarma-disorder}. In an ultraclean graphene sample, the conductance can be suppressed well below $e^2/h$ if the average charge density is sufficiently reduced near the neutrality point. In other words, it is the saturated carrier density around the Dirac point due to the presence of electron-hole puddles that determines the minimum of conductance in graphene. Using the modified Drude model for mobility which includes the impurity-induced effective charge concentration, we can estimate a density-independent mobility and also the contact resistance via the following equation for total resistance in a two-probe measurement~\cite{modifiedDrudeModel}:
	
	\begin{equation}
	\begin{split}
	R_{total} & = R_{contact} + R_{graphene} \\&
	 = R_{contact} + \frac{(L/W)}{e\mu \sqrt{{\delta n}^2 + n_{gate}^2}},
	\end{split}
	\end{equation}
	
\noindent where, $L$ and $W$ are the channel length and width, respectively, $\mu$ is the mobility, and $n_{total}\equiv \sqrt{{\delta n}^2 + n_{gate}^2}$ is the total charge density determined by the residual concentration $\delta n$ due to impurities and the gate-modulated density $n_{gate}$. By fitting this equation to resistance curve vs gate-induced charge density, we estimated the mobility and the density fluctuation $\delta n$. A density inhomogenity of $\delta n = 4 \times 10^9$ cm$^{-2}$ is obtained which is consistent with the full width at half maximum of the Dirac peak. The very low residual charge density implies an ultraclean sample with extremely low impurities. The electron mobility for the cleaned sample is estimated as $1.2 \times 10^5$ cm$^2$V$^{-1}$s$^{-1}$ which is among the highest mobilities achieved for suspended graphene devices~\cite{9,SuspendedUltrahighM,12,13,14}. Although it is hard to extract mean free path without knowing the precise contact resistance value, using the semiclassical relation between mobility and mean free path~\cite{6} $\sigma = ne\mu = \frac{2e^2}{h}(\sqrt{\pi n}~l_e)$, we can roughly estimate $l_e\sim 0.6~\mu$m at a density $n = 2\times 10^{11}$ cm$^{-2}$. On the other hand, we note that the mean free path and therefore the mobility in two-probe geometry is limited by the separation of the probes that is the length of the channel $L$. In the ballistic limit, the mean free path has a maximum of $\sim L/2$ bound by the boundary conditions imposed by the two-lead configuration, and it is almost independent of the carrier density except at the neutrality point. As a result, one would measure the device mobility rather than the intrinsic material mobility in short devices with two-probe geometry~\cite{12}. The mean free path we obtained above is consistent with the ballistic transport. Consequently, the ballistic mobility scales with $n^{-1/2}$ diverging at lower densities where it reaches $\sim 2.5 \times 10^5$~cm$^2$V$^{-1}$s$^{-1}$ at $n = 4\times 10^{9}$ cm$^{-2}$.%, which means only a few scatterers remained over the graphene surface. 

The sample is then taken through an in-situ thermo-pressure cycle from 1.5 K to 200 K along with loosening of the vacuum ($\sim 1\times10^{-6}$ mBar) up to $10^{-2}$~mBar and then cooled back to cryogenic temperatures after which it adopted strong conductance fluctuations leading to an insulating behavior around the charge neutrality point with mega-ohm resistance peaks. We believe that the ultraclean sample was disordered during the thermo-pressure cycle by some adsorbents accompanying strong short-range potentials leading to pronounced conductance fluctuations and intervalley backscattering~\cite{suzuura2002crossover,suzuura_anderson}. The conductance exhibits strong fluctuations as the  charge density is varied and an insulating behavior at low density regime. In Fig.~\ref{G-V-T}(a), the conductance as a function of gate voltage is plotted at various temperatures up to 30~K. The conductance fluctuations are strongly dependent on temperature especially around the neutrality point and are remarkably suppressed at higher temperatures. A comparison between the gate-dependent conductance before and after the TPC is illustrated in Fig.~\ref{G-V-T}(b). The adsorption of atomic impurities during the TPC caused a suppression of conductance along with strong fluctuations around the Dirac point. It is also noted that the sample mobility degraded by almost an order of magnitude to $1.5 \times 10^4$ cm$^2$V$^{-1}$s$^{-1}$. This corresponds to an elastic mean free path of $l_e\sim 0.1~\mu$m. 

The decrease of mobility and mean free path after the TPC clearly indicates that the sample has acquired excess charge scatterers during the TPC.
It is known that some type of adsorbates, impurities, vacancies, or defects can induce strong resonant scatterers that significantly limit the mobility of the graphene devices~\cite{Resonance2,Resonance1,Resonance3,Resonance4,Resonance5,Resonance6,Resonance7,Resonance8}
For example, physisorbed oxygen molecules are shown to form resonant states above the Dirac point and to decrease the electron mobility. Only in a non-suspended graphene sheet on SiO$_2$ substrate, O$_2$ molecules may interact with SiO$_2$ at the interface of the Graphene and the substrate which would result in a hole doping. But in the case of free-standing graphene, the oxygen molecules adsorbed on a clean graphene surface do not transfer the charge and therefore causes negligible doping~\cite{Resonance9}. The reduction of the electron mobility in our suspended sample after TPC is indeed consistent with the presence of resonant scatterers.
Since the sample was undergone the TPC inside the chamber of the dilution refrigerator, the most likely impurities are physically adsorbed water and oxygen molecules. They can make relatively strong bonds with carbon atoms in graphene via van der Waals interaction with adsorption energies of about 100 meV~\cite{Adsorbent1,Adsorbent2,Adsorbent3}. 
These adsorbents produce sharp potentials in atomic scales which may act as a source of intervalley backscattering. Further studies with controlled physisorption of cleaned graphene samples are needed for complete understanding of the effect of physical adsorption on graphene's transport properties.

\begin{figure}[t]
\begin{center}
{{\includegraphics[width=0.5\textwidth]{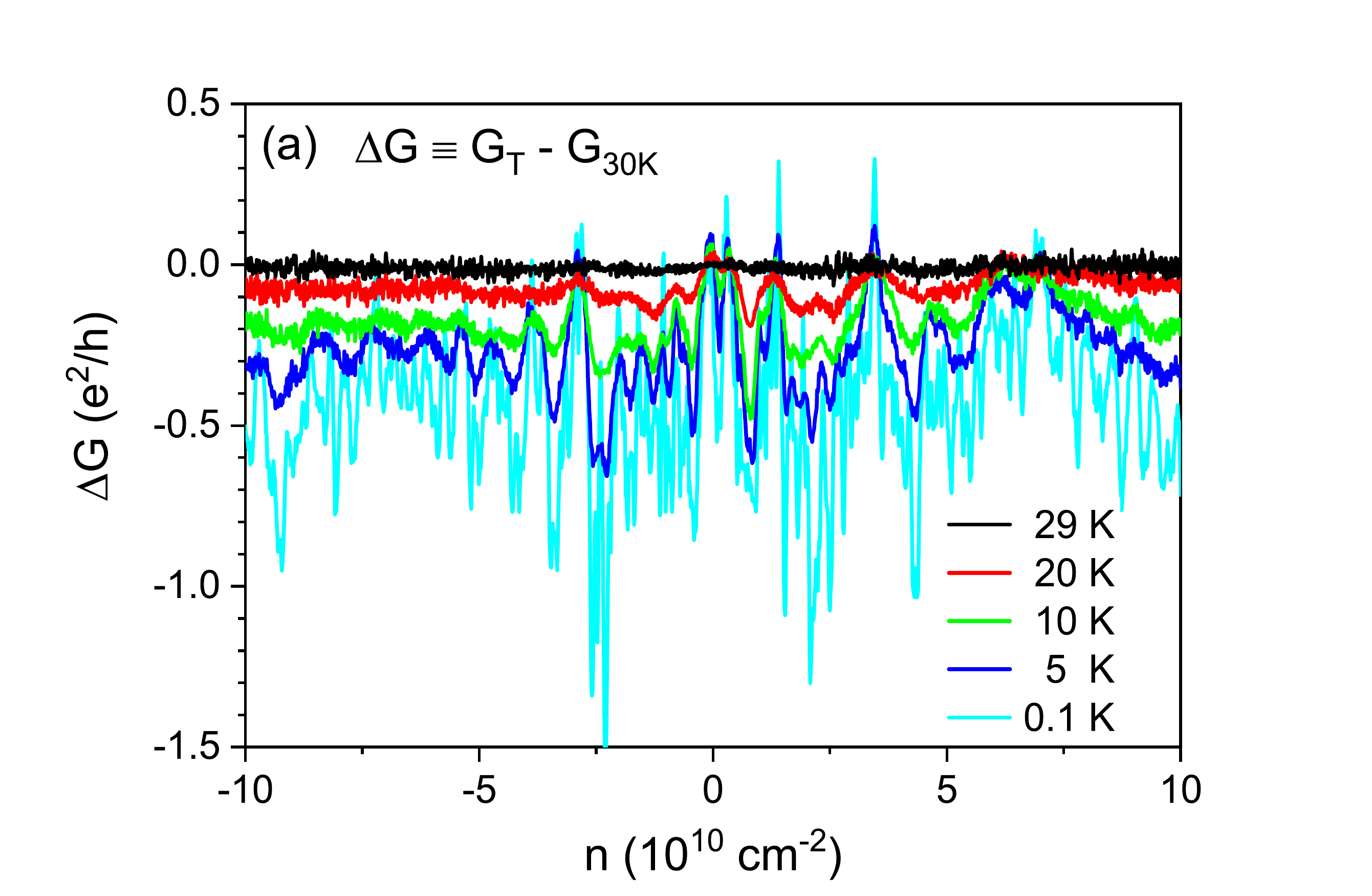} }}
{{\includegraphics[width=0.5\textwidth]{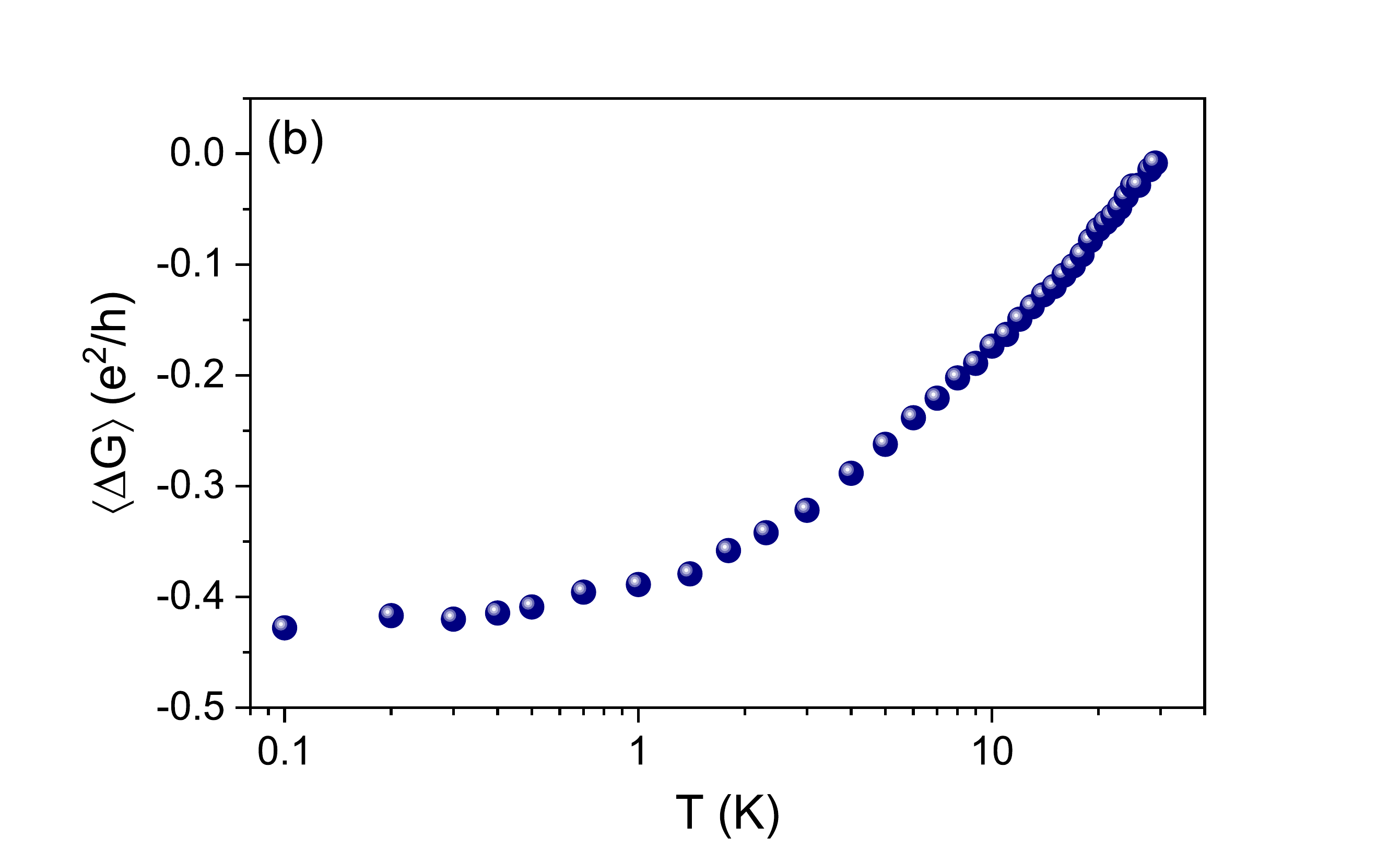} }}
\caption{(a) Change of the conductance, $\Delta$G relative to its value at 30~K with respect to the carrier density, $n$. The conductance fluctuations as well as the average change are strongly temperature dependent. (b) $\Delta$G averaged over the density window depicted in (a) shows a logarithmic behavior at high temperatures and tends to saturate for T~$\le 2$~K where the coherence length of the electrons exceeds the sample size.}\label{DeltaG}
\end{center}
\end{figure}

As the temperature is lowered, the transport of electrons becomes coherent and leads to quantum interference corrections to the conductance. Depending on the nature of disorder, graphene may exhibit weak localization or anti-localization behaviors. In low mobility devices, for example graphene on silicon oxide substrate, where the long-range impurities with spatially slowly varying potentials are the dominant scatterers, the sublattice symmetry is preserved therefore intervalley backscattering is prohibited and a weak anti-localization behavior is expected~\cite{suzuura2002crossover,mccann2006weak}. However, even in the absence of intervalley scattering, some types of long-range disorders may effectively break the time reversal symmetry leading to the suppression of low field magnetoresistance~\cite{Long_range_disorder3}. On the other hand, in ultraclean graphene samples, the short-range scattering dominates resulting in negative quantum interference corrections to the conductivity. In Fig.~\ref{DeltaG}(a), the relative fluctuations of conductance is illustrated in low density regime at different temperatures. It can be seen that as the temperature is decreased the fluctuations in the conductance are strongly pronounced especially around the Dirac point generating the insulating dips seen in Fig.~\ref{G-V-T}(a). The fluctuations are reproducible at different temperatures while intensifying at lower temperatures such that they can  diminish the conductance occasionally around the Dirac point and lead to an insulating behavior when the carriers are totally localized in the bulk. Suzuura and Ando~\cite{suzuura2002crossover} showed that in two-dimensional honeycomb lattice the quantum interference correction to the Boltzmann conductivity is given by $\Delta \sigma = \pm (e^2/\pi h)$~log$(l_\phi/l_{e})$, where $l_\phi$ and $l_{e}$ are the coherence and elastic scattering lengths, respectively. In the case of long-range disorders, the backscattering is forbidden thus the correction is {\em positive} whereas in the presence of short-range potentials the intervalley scattering becomes probable and lead to a {\em negative} correction to conductivity. The coherence length decreases at higher temperatures, suppress the quantum interference effects and lead to a logarithmic temperature behavior for the conductance correction. The average value of change in the conductance at different temperatures is plotted in Fig.~\ref{DeltaG}(b) which shows a logarithmic suppression of the negative conductance correction for T~$\geq 2$~K. Below 2~K, the conductance correction starts to saturate. This is the temperature below which the coherence length exceeds the sample size thus saturates the conductance.  

\begin{figure}[t]
	\begin{center}
		\includegraphics[width=0.5\textwidth]{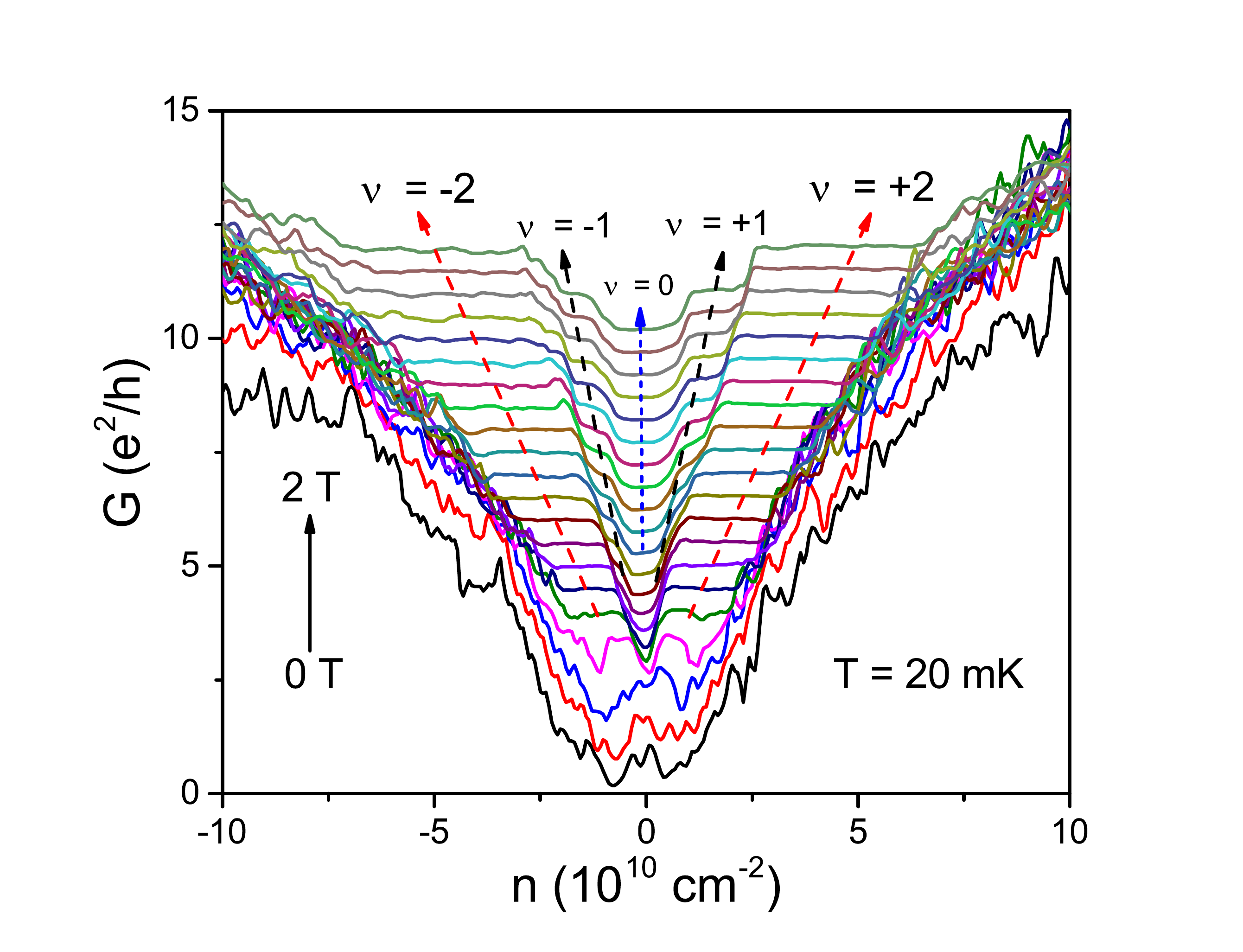}
		\caption{Conductance of the suspended graphene sample after it is undergone TPC. Plots are taken at magnetic fields between $B=0$ and $2~T$ in $0.1~T$ steps. Dips of conductance where sample is insulating  gradually fade with increasing B and quantum Hall plateaus for filling factors $\nu=\pm2$ form. $\nu=0,\pm1$ plateaus also start to appear at around $1~T$. The plots are offset by a constant amount.}\label{G-n-B}
	\end{center}
\end{figure}

Since the insulating behavior is observed for $\lvert n\lvert \le 10^{11}$~cm$^{-2}$ and the localization requires a mean free path of the order of the Fermi wavelength, $\lambda_F=(4\pi/n)^{1/2} $, we can estimate the mean free path for intervalley scattering as $l_{iv}\sim 0.1~\mu$m. A similar length scale is also inferred from the magnetotransport data presented in Fig.~\ref{G-n-B}. The field at which the sample transitions to the quantum Hall state $(B\sim 0.4~T)$ gives a length scale $(\phi_0/B)^{1/2}\approx0.1~\mu$m corresponding to a flux quantum $\phi_0 = h/e$ enclosed by cyclotron orbits which sets a minimum on the intervalley scattering length. Note that the intervalley scattering lengths estimated from localization behavior and magnetoresistance measurements are in agreement with the mean free path we obtained from mobility in the previous section.

The intervalley scattering is also manifested in the quantum Hall regime. Fig.~\ref{G-n-B} shows the conductance as a function of carrier density at various magnetic fields from 0 to 2 tesla. A direct transition from the insulating behavior around the Dirac point to quantum Hall regime is observed around $0.4~T$ where a single conductance minimum at Dirac point appears with the development of $\nu=\pm2$ plateaus around it. 
%This behavior indicates that the bulk conductance is continuously vanishing with the current transport through edge channels until it fully enters the quantum Hall regime. 
Moreover, the sample displayed clear $\nu=0,\pm1$ plateaus besides the normal sequence of plateaus for a single layer graphene at relatively low magnetic fields. The presence of inter-valley scattering lifts the valley degeneracy and splitting the spin degeneracy at sufficiently high magnetic fields ($\gtrsim1$~T) giving rise to the fully symmetry-broken quantum Hall sequences~\cite{DasSarmaRevModPhy,13} which can be resolved only in ultra-clean samples with small amount of short-range disorders. 

The zeroth Landau level (LL) has an anomalous structure different from other LL's. It was shown by Ref.~\cite{SymBrk} that for non-zero LL's the ground states at half filling i.e., $\nu = \pm 4, \pm 8,...$ are spin polarized due to dominant Zeeman splitting while quarter filling states are valley polarized. The situation is reversed for zeroth LL, where the $\nu=0$ state, which corresponds to half filling of the zero energy LL, is unpolarized and spin textured excitations form at fully polarized $\nu = \pm1$ states. Therefore, in our sample, the presence of $\nu=0$ plateaus upon the formation of LL's indicates the sublattice symmetry breaking which can only be explained by atomically sharp potentials on a suspended graphene. At higher magnetic fields, Zeeman splitting of the zero LL leads to $\nu=\pm1$ states.

%\begin{figure}[t]
%\begin{center}
%\includegraphics[width=0.45\textwidth]{Fig4_V1}
%\caption{\label{Fig4} Zero-field conductance fluctuations root-mean-square as a function of temperature after thermal cycle.}
%\end{center}
%\end{figure}

%\begin{figure}[t]
%\begin{center}
%\includegraphics[width=0.45\textwidth]{FFig4_V1}
%\caption{\label{Fig5} Two-terminal conductance at different magnetic fields up to $2~T$. Insulating peaks declines monotonously to QH $\nu=\pm2$ plateaus. The $\nu=0,\pm1$ plateaus also appear at around $1~T$. The graphs are stacked with a constant amount.}
%\end{center}
%\end{figure}

%\begin{figure}[t]
%\begin{center}
%\includegraphics[width=0.45\textwidth]{Fig6_V1}
%\caption{\label{Fig6} Two-terminal conductance in quantum Hall regime measured at base temperature $T=~20mK$ compared between before and after thermal cycle. }
%\end{center}
%\end{figure}

In this paper, we have investigated an unexpectedly low conductance at the Dirac point of a current-annealed micron-size suspended graphene sheet, well below the Boltzmann conductivity for graphene, $e^2/h$. More interesting observation was a highly temperature-dependent insulating behavior in the suspended device after being disordered by sharp atomic-scale impurity potentials during a thermo-pressure cycle. Such a low conductance around the charge neutrality point well below the ballistic limit $4e^2/\pi h$, before and after thermal cycle, indicates that the short-range intervalley scatterers dominated over the long-range disorders. This behavior arises from the suppression of the potential inhomogeneities induced by charge puddles near the neutrality point of high quality graphene samples, which may incorporate a vanishing conductance and metal-insulator transition~\cite{16,Amet_PRL2013,DasSarmaRevModPhy}. The adsorbent-induced intervalley scattering brought the sample into a completely insulating regime near Dirac point. Lifting of the valley symmetry due to strong inter-valley scattering was also reflected in the quantum Hall measurements as the $\nu=0$ plateaus appeared at relatively small fields of $\sim 0.5~T$.

The authors would like to thank Inanc Adagideli for his fruitful discussions.
This work is funded by Scientific and Technological Research Council of Turkey (TUBITAK) under Project Grant No. 112T990.

\bibliographystyle{apsrev4-1}
\bibliography{StrongLocalization}
\end{document}